\documentclass[12pt]{article}

\usepackage{amsthm,amsmath,amsfonts,algorithm,graphicx, physics}
\usepackage{fullpage,authblk}

\theoremstyle{definition} 
\theoremstyle{definition} 
\newtheorem {theorem} {Theorem}

\newcommand{\kb}[1]{\ketbra{#1}}
\newcommand{\bk}[1]{\braket{#1}}

\newcommand{\quart}{\frac{1}{4}}
\newcommand{\half}{\frac{1}{2}}

\newcommand{\pib}{p_{ib}}
\newcommand{\pjb}{p_{jb}}
\newcommand{\pphib}{p_{\phi b}}
\newcommand{\pphij}{p_{\phi j}}

\newcommand{\pphii}{p_{\phi i}}

\newcommand{\pphiphi}{p_{\phi \phi}}

\newcommand{\pii}{p_{ii}}
\newcommand{\pij}{p_{ij}}
\newcommand{\pji}{p_{ji}}
\newcommand{\pjj}{p_{jj}}

\newcommand{\bkfifj}{\bk{f_i}{f_j}}

\newcommand{\fj}{f_j}

\newcommand{\smbni}{\sum_{b \ne i}}

\newcommand{\piphi}{p_{i \phi}}
\newcommand{\pjphi}{p_{j \phi}}

\newcommand{\ebi}{e_b^i}
\newcommand{\ebj}{e_b^j}
\newcommand{\fb}{f_b}

\newcommand{\ot}{\otimes}

\newcommand{\mbi}{\mathbb{I}}

\newcommand{\eii}{e_i^i}
\newcommand{\eij}{e_i^j}
\newcommand{\eji}{e_j^i}
\newcommand{\ejj}{e_j^j}

\newcommand{\kebj}{\ket{e_b^j}}

\newcommand{\kbphi}{\ketbra{\phi}}

\newcommand{\kbi}{\ketbra{i}}
\newcommand{\kbj}{\ketbra{j}}

\newcommand{\ki}{\ket{i}}

\newcommand{\kj}{\ket{j}}

\newcommand{\kphi}{\ket{\phi}}

\newcommand{\keii}{\ket{e_i^i}}

\newcommand{\keji}{\ket{e_j^i}}

\newcommand{\kfi}{\ket{f_i}}
\newcommand{\kfj}{\ket{f_j}}
\newcommand{\bfi}{\bra{f_i}}
\newcommand{\bfj}{\bra{f_j}}

\newcommand{\kf}{\ket{f_b}}

\newcommand{\slim}{b \ne i, b \ne j}

\newcommand{\sqtt}{\frac{1}{\sqrt{2}}}

\floatname{algorithm}{Protocol}

\begin{document}
	\title{Analysis of a High-Dimensional Extended B92 Protocol}

\author[1]{Hasan Iqbal\footnote{Email: \texttt{hasan.iqbal@uconn.edu}}}
\author[1]{Walter O. Krawec}
\affil[1]{\small{Department of Computer Science and Engineering}\\\small{University of Connecticut}\\\small{Storrs, CT 06269 USA}}
	
\date{}
	
	\maketitle
	
	\begin{abstract}
		Quantum key distribution (QKD) allows two parties to establish a shared secret key that is secure against all-powerful adversaries.  One such protocol named B92 is quite appealing due to its simplicity but is highly sensitive to channel noise.  In this work, we investigate a high-dimensional variant of an extended version of the B92 protocol and show that it can distill a key over high noise channels. The protocol we consider requires that Alice send only three high-dimensional states and Bob only perform partial measurements.  We perform an information-theoretic security analysis of our protocol and compare its key rate to that of a high-dimensional BB84 protocol over depolarization and amplitude damping channels. 
	\end{abstract}

	\section{Introduction}
	The need for perfect security necessitated the development of cryptographic systems where there are no computational constraints on the capabilities of the adversary.  Quantum key distribution (QKD) is one such system that is extensively studied and is increasingly maturing to the point of real-world adoption. In QKD, using quantum mechanical properties of communication resources, two parties Alice (A) and Bob (B), following a specified set of steps, generate a shared secret that is secure from an all-powerful adversary Eve (E). 
	% brief history of QKD
	
	Since the first QKD protocol by Bennett and Brassard in $1984$ (BB84) \cite{QKD-BB84}, there have been numerous advances in both theoretical and practical aspects \cite{qkd-survey-pirandola, qkd-survey-scarani, qkd-survey-akshata}. However, because generating, maintaining, and manipulating quantum resources are exceptionally hard with current technologies, people have strived to create conceptually simpler protocols that also require less quantum resources. For instance, BB84 itself uses four quantum states and two measurement bases. In 1992, Bennett proposed an even simpler QKD protocol called B92, that uses only two non-orthogonal states and measurement bases \cite{bennett1992quantum}.

	The unconditional security of this protocol has been investigated by several authors \cite{tamaki2003unconditionally,matsumoto2013improved} with continually improving results (for instance, in \cite{matsumoto2013improved}, a noise tolerance of $6.5\%$ is reported). However, B92 is very noise sensitive compared to other protocols like BB84 as was already noted in the original paper \cite{bennett1992quantum}.  Lucamarini et al. \cite{lucamarini2009robust}, proposed an extended version of B92 (Ext-B92) which added two additional non-informative states to better bound Eve's information gain.  Depending on the user-defined choice for key encoding states, the noise tolerance of that protocol can approach $11\%$ in the asymptotic scenario \cite{lucamarini2009robust}, similar to BB84, and at least $7\%$ in the finite key scenario \cite{amer2020finite}.

	These protocols mentioned above use qubits (dimension two systems) as the communication resource between Alice and Bob. However, higher dimensional quantum systems (see \cite{hd-survey} for a brief survey) have been shown to have several advantages and interesting properties over qubit-based protocols.  Some protocols have been shown to withstand a high channel noise level as the dimension of the system increases \cite{bechmann2000quantum,chau2005unconditionally,sheridan2010security,vlachou2018quantum}.  Others exhibit interesting theoretical properties such as the so-called ``Round Robin'' protocol which can bound Eve's information based only on the dimension of the system and not necessarily through observing channel noise \cite{sasaki2014practical}.  In addition to several theoretical results that prove the unconditional security of HD-QKD protocols, the actual technology to implement high-dimensional systems is also becoming more mature with recent high-dimensional protocols proving to be feasible to implement \cite{wang2020high, islam2017provably, mower2013high, da2021path, lee2019large}. Thus, it is worth studying protocols that are highly susceptible to noise, like B92 based variants (in our case the extended B92), to see if HD-systems give an advantage.

	In this work, we propose a high-dimensional variant of the Ext-B92 protocol of \cite{lucamarini2009robust}.  Keeping in mind that certain high-dimensional states are difficult to create or distinguish, we make sure in our protocol to limit the required state preparations and measurement operations required.  In particular, Alice need only be able to send three high dimensional states while Bob need only be able to perform a computational basis measurement (distinguishing any computational state $\{\ket{0}, \ket{1}, \cdots, \ket{D-1}\}$) or be able to perform a partial basis measurement in an alternative basis - this partial measurement need only distinguish a particular superposition state defined in the protocol and need not be capable of distinguishing all $D$ possible states.  As far as we are aware, this is a novel high-dimensional QKD protocol.
	
	%          In particular, we only require parties to prepare and distinguish computational states $\{\ket{0},\cdots, \ket{D-1}\}$ along with only one distinguished superposition state.  That is, we do not require users to be able to, say, prepare and measure any other state besides these $D+1$; nor do we require users be able to perform a complete measurement in two distinct bases.  Furthermore, independent of $D$, we require only that Alice be able to prepare three distinct states: two computational ones, and on superposition state.
	Despite these limitations on Alice and Bob's capabilities, we show that these higher dimensional states do help improve noise tolerance in this protocol as the dimension of the system increases which agrees with recent research on high-dimensional BB84. We perform an information-theoretic security analysis and show that it can maintain a positive key rate while withstanding noise levels of $5.35\%$ for qubits (dimension $2$) to $15.5\%$ for dimension $2^{14}$ in a depolarizing channel. Thus, we show that higher dimensions can aid in depolarization noise tolerance for a B92 style encoding scheme with (partial) extended test cases in the form of a third basis state being transmitted for testing the channel.  Moreover, we consider an amplitude damping channel and show that choosing the distinguished superposition state carefully in a high-dimensional QKD protocol is of significant importance as different choices lead to different noise tolerances. 
	
	We make several contributions in this work.  First, we describe and analyze a high-dimensional version of the Ext-B92 protocol originally introduced for qubits \cite{lucamarini2009robust}.  We perform an information theoretic security analysis against collective attacks (a powerful class of attacks against QKD protocols) to derive its key-rate in the asymptotic scenario for arbitrary dimensions and channel parameters.  Our methods here may have a broader impact in other QKD protocol security analyses, especially for high-dimensional systems with only partial basis measurements or state preparations as with our protocol here.  Finally, we evaluate our resulting key rate and compare it with a high-dimensional version of the BB84 protocol, showing how the choice of states to send can greatly affect the key-rate depending on the channel.
	
	\section{Notation}
	For a quantum system $A$ we will use $\rho_A$ to denote its density operator. Its von Neumann entropy will be denoted by $S(A) = S(\rho_A)$ and is defined by $-\tr(\rho_A \log \rho_A)$. Given a bipartite quantum state $\rho_{AE}$ shared by two systems $A$ and $E$, we will denote the conditional von Neumann entropy of $A$ given access to $E$, by $S(A|E)_\rho$. We will often forgo writing the subscript $\rho$ when the context is clear. This conditional entropy is defined as $S(AE) - S(E)$.  The Shannon entropy of $A$ will be denoted by $H(A)$ and the conditional Shannon entropy of two systems $A$ and $B$, will be denoted by $H(A|B)$. The binary entropy function will also be represented by $H(p)$ where $H(p) = -p \log(p) - (1 - p) \log(1 - p)$ for $p \in [0, 1]$. All logarithms presented in this work  are base $2$.  For an arbitrary quantum state $\ket{\psi}$, we use $P(\ket{\psi})$ to denote its projector $\kb{\psi}$. Finally, given a vector $\ket{x}$ and a numerical value such as $\frac{1}{2}$ we sometimes write $\ket{\frac{1}{2}x}$ to mean $\frac{1}{2}\ket{x}$. %%To denote the real part of an inner product $\bk{a}{b}$ between two quantum states $\ket{a}$ and $\ket{b}$, we will use $\Re\bk{a}{b}$. The tensor product between $\ket{a}$ and $\ket{b}$ will be denoted by both $\ket{a, b}$ and $\ket{a} \otimes \ket{b}$.
	
	Later, to compute the lower bound of the conditional von Neumann entropy of a classical-quantum state $\rho_{AE}$, we make use of the following theorem:
	
	\begin{theorem}\label{theorem1} (From \cite{QKD-Tom-Krawec-Arbitrary-2016}): 
		Let $\mathcal{H}_A\otimes \mathcal{H}_E$ be a finite dimensional Hilbert space. Consider the following density operator.
		\begin{align}
			\rho_{AE} = \frac{1}{N}\left(\kb{0}_{A}\otimes\sum_{i=1}^{\tau}\kb{e_{i}^{0}}+\kb{1}_{A}\otimes \sum_{i=1}^{\tau}\kb{e_{i}^{1}}\right) , \nonumber
		\end{align}
		where $N>0$ is a normalization term, $\tau<\infty$, and each $\ket{e_{i}^{j}}\in \mathcal{H}_E$ (these are not necessarily normalized, nor orthogonal, states; also it might be that $\ket{e_{i}^{j}}\equiv 0$ for some $i$ and $j$). Let $n_{i}^{j} = \bk{e_{i}^{j}}\geq 0$. Then:
		\begin{align}
			S(A|E)_{\rho} \geq \sum_{i=1}^{\tau}\left(\frac{n_{i}^{0}+n_{i}^{1}}{N}\right)\cdot S_{i}, \nonumber
		\end{align}
		where:
		\begin{align}
			S_{i} =
			\begin{cases}
				h\Big(\frac{n_i^{0}}{n_i^{0} + n_i^{1}}\Big) - h(\lambda_i)       & \quad \text{if } n_i^{0} > 0 \text{ and } n_i^{1} > 0,\\
				0  & \quad \text{otherwise.} \nonumber
			\end{cases}
		\end{align}
		and:
		\begin{align}
			\lambda_i = \frac{1}{2}+\frac{\sqrt{(n_i^0-n_i^1)^2 + 4 \Re^{2}\braket{e_i^0}{e_i^1}}}{2(n_i^0+n_i^1)}. \nonumber
		\end{align}
	\end{theorem}
	
	\section{The Protocol}
	The protocol we propose here is a high-dimensional variant of the Ext-B92 protocol originally described in \cite{lucamarini2009robust}.  In that protocol, two non-orthogonal states, similar to B92, are used for key encoding while two additional states are used for quantum tomography (these four states together come from two distinct bases).  In the higher dimensional case we analyze here, we will use two non-orthogonal states for key encoding; for error testing, we adopt a simplification from \cite{amer2020finite} (done there for the qubit case) and not require users to be able to control two complete bases.  More specifically, in our high-dimensional extended B92,  Alice sends $\ket{i}$ and $\kphi = \sqtt (\ki + \kj)$ to encode classical key bits of $0$ and $1$ respectively, where $\ki, \kj$ are fixed and chosen from the $D$-dimensional computational basis states $\{\ket{0}, ... , \ket{D - 1}\}$. We ask that Alice send only $\kj$ as the additional uninformative state.  Thus, Alice need only be able to prepare and send three distinct quantum states. On Bob's part, we require his ability to measure in two POVMs. These are $Z = \{ \kb{0}, ... , \kb{D - 1} \}$ (the complete computational basis) and $X = \{ \kbphi, \mbi - \kbphi \}$ where of course, the identity operator $\mbi$ is understood to be $D$-dimensional. Hence, Bob would be able to detect any computational basis state but would only need to detect $\kphi$. Our protocol, which we call here HD-Ext-B92, in detail appears in Protocol \ref{prot:hd-b92}.

	\begin{algorithm}
		\caption{High-dimensional Extended B92 (HD-Ext-B92) \label{prot:hd-b92}}
		
		$ $\newline
		\textbf{Public Parameters:} The dimension of a signal state $D \ge 2$ and the choice of distinct $i, j \in \{0,1,\cdots, D-1\}$ are arbitrary, but are fixed at the start of the protocol and known to all parties (including the adversary).
		$ $\newline
		\textbf{Quantum Communication Stage:} The quantum communication stage of the protocol will repeat the following until a sufficiently large raw-key has been distilled:
		\begin{enumerate}
			\item Alice chooses randomly whether this round will be a ``key-round'', where a raw key bit will attempt to be established, or a ``test'' round, which will be used for error testing later.  If this is a key-round, she will choose a random key bit and if this is $0$, she will prepare and send the state $\ki$; otherwise, she sends the state $\kphi$.  If this is a test round, she will prepare $\ki$, $\kj$, or $\kphi$ choosing uniformly at random.
			
			%			Alice flips an unbiased coin to choose her classical bit, whether it should be a $0$ or a $1$ CONFIRMGRAMMAR. She sends an $\ki$ if her classical bit is a $0$. For classical bit $1$, she sends a $\kphi$. We denote these rounds, where a key-bit can be shared between Alice and Bob, as key-rounds. On the other hand, she can decide a particular round to be a test round. Where, she would send any of $\ki, \kj$ or $\kphi$, but these rounds would not count towards the generated key. These rounds can be used for channel parameter estimation. 
			\item Bob measures in either the $Z$ basis or using POVM $X$. In a key-round, if he uses $Z$ and observes any outcome other than $\ki$, then he sets his bit to be 1. Otherwise, if he uses POVM $X$ and observes $\mbi - \kbphi$, then he sets his bit to be 0.  All other results are considered inconclusive.
			
			\item Alice informs Bob over the authenticated channel whether this was a test round or a key-round.  If this is a key-round, Bob also tells Alice if his result was inconclusive (in which case both parties discard the iteration).  On test-rounds, both parties disclose their choices and measurement outcomes to determine the error rate in the channel.  In particular, they will observe those statistics enumerated in Table \ref{tab:observable_stats}.  Note that we will not discard mismatched basis events; i.e., events where Alice and Bob use different bases.  Indeed, such events can greatly improve key generation rates \cite{barnett1993eavesdropping,watanabe2008tomography,matsumoto2008key,krawec2016asymptotic,tamaki2014loss} and so we use this technique here.
		\end{enumerate}
		\textbf{Classical Communication Stage:} 
		Alice and Bob will next run an error correction protocol and a privacy amplification protocol resulting in a secret key of size $\ell$ bits (possibly $\ell = 0$ if it is determined that Eve has too much information, to be discussed later in this paper, and so parties abort the protocol).
	\end{algorithm}

	\begin{table}[ht]
		\centering
		\caption{Definition of Alice and Bob's directly observable parameters ($\ket{b} \in \{\ket{0}, ... , \ket{D - 1}\}$)}
		\begin{tabular}[t]{|c|l|}
			\hline
			Parameter  &  Description of Probability Value\\
			\hline
			$p_{ib}$ &  Bob observes $\ket{b}$ conditioned on Alice sending $\ki$ and Bob choosing the $Z$ basis\\
			$p_{jb}$ & Bob observes $\ket{b}$ conditioned on Alice sending $\kj$ and Bob choosing the $Z$ basis\\
			$p_{\phi b}$ & Bob observes $\ket{b}$ conditioned on Alice sending $\ket{\phi}$ and Bob choosing the $Z$ basis\\
			$p_{i \phi}$ & Bob observes $\ket{\phi}$ conditioned on Alice sending $\ki$ and Bob choosing POVM $X$ \\
			$p_{j \phi}$ & Bob observes $\ket{\phi}$ conditioned on Alice sending $\kj$ and Bob choosing POVM $X$ \\
			$p_{\phi \phi}$ & Bob observes $\ket{\phi}$ conditioned on Alice sending $\ket{\phi}$ and Bob choosing POVM $X$ \\								
			\hline
		\end{tabular}
		\label{tab:observable_stats}
	\end{table}

	\section{Security Analysis}

	%% Begin with entropy computation, which is based on inner products which leads to next section...	
	
	In the quantum communication stage of our protocol, Alice and Bob use the quantum channel to establish a raw-key. Because Eve has total control over this channel, she may attack the traveling signals arbitrarily while only respecting the laws of physics. In this paper, we consider collective attacks whereby Eve attacks each round of the protocol independently and identically, but may delay her measurements until the end of the protocol. These are a powerful class of attack which often imply security of general coherent attacks \cite{konig2005finetti,christandl2009postselection}, though we leave a complete proof of whether this applies to our protocol as future work.
	
	The goal of our analysis is to obtain a lower bound on the conditional von Neumann entropy $S(A|E)$, which represents how much entropy is left in Alice's register $A$, given Eve's (quantum) memory $E$. Then, we will find how much this quantity differs from the conditional Shannon entropy $H(A|B)$, which represents how much entropy is left in Alice's register given Bob's memory $B$. These two terms will ultimately let us calculate our quantity of interest from this protocol, which is, the key rate $r$ (namely, the number of secret key bits, denoted $\ell$ over the size of the produced raw key denoted $M$). To compute the key rate we use the Devetak Winter key-rate \cite{devetak2005distillation, renner2005information}, which states the key rate $r$ in the asymptotic setting is:
	\begin{align}
		r = \underset{M \rightarrow \infty}{\lim} \frac{\ell}{M} = \inf[S(A|E) - H(A|B)], \label{eq:devetakwinter}
	\end{align}
	where the infimum is over all collective attacks performed by Eve that fall within the range of observed noise statistics (in our case, those statistics shown in Table \ref{tab:observable_stats}).  Note that the above entropy functions are computed over a single key-round. However $S(A|E)$ is not straightforward to calculate, unlike $H(A|B)$, because it involves Eve's quantum memory on which we only have partial information. Nevertheless, we can obtain a lower bound on $S(A|E)$ which will be the main goal of our security analysis.
	
	We begin by modeling the state of the joint quantum system held between Alice, Bob, and Eve at the end of one key-round of the protocol.  That is, to compute Equation \ref{eq:devetakwinter}, we need the von Neumann entropy of the resulting density operator conditioned on a key bit being distilled and so we must model the joint quantum state, conditioning on the event that Alice and Bob establish a key bit.
	
	At the beginning of the protocol, Alice decides on her classical bit and sends her qudit accordingly to Bob. If she wants to send classical bit $0$, she sends a $\ki$ and if she wants to send $1$, she sends a $\kphi$. So when she sends the qudits, her own classical register, denoted by $A$ and the transit register, denoted by $T$ (used to model the traveling qudit), is in the following state:
	\begin{align*}
		\rho^{(0)}_{AT} = \half \kb{0}_A \ot \kbi_T + \half \kb{1}_A \ot \kbphi_T.
	\end{align*}
	Eve attacks this traveling qudit with a unitary attack operator $U$, which acts on Hilbert space $\mathcal{H}_T \ot \mathcal{H}_E$. Here, $\mathcal{H}_E$ models Eve's memory space. Assuming her own memory is in an arbitrary but normalized state $\ket{\chi}_E$ that resides in $\mathcal{H}_E$, we can describe $U$'s action on a basis state $\ket{a}$ as: 
	\begin{align*}
		U\ket{a}_T \ot \ket{\chi}_E = \sum_{b = 0}^{D - 1} \ket{b, e_b^a}_{TE},
	\end{align*}
	where $D$ is the dimension of each qudit and each $\ket{e_b^a}$ is an arbitrary state in Eve's ancilla. Because $U$ is unitary, we note that the following must hold:
	%	\begin{align*}
	%	1 = \bra{i} U^\dagger U \ki &= \Big(\sum_{b = 0}^{D - 1} \bra{b, \ebi}_{TE}\Big) \Big(\sum_{c = 0}^{D - 1} \ket{c, \ebi}_{TE}\Big) = \sum_{b, c = 0}^{D - 1} \bk{b}{c} \bk{\ebi}{e_c^i} = \sum_{b = 0}^{D - 1} \bk{\ebi}{e_b^i}.
	%	\end{align*}
	%\begin{align*}
	$\sum_{b = 0}^{D - 1} \bk{\ebi}{e_b^i} = 1.$
	%\end{align*}
	Additionally, by linearity of $U$ we have:
	\begin{align}
		U \kphi_T &= U \sqtt (\ki_T + \kj_T) \nonumber\\
		%	&= \sqtt (U\ki_T + U \kj_T) \nonumber\\
		%	&= \sqtt \big(\sum_{b = 0}^{D - 1} \ket{b, \ebi}_{TE} + \sum_{b = 0}^{D - 1} \ket{b, \ebj}_{TE} \big) \nonumber \\
		&=  \sqtt  \sum_{b = 0}^{D - 1} \ket{b}_T \ot (\ket{\ebi}_E + \ket{\ebj}_E) \nonumber\\
		&=  \sqtt \sum_{b = 0}^{D - 1} \ket{b}_T \ot \ket{f_b}_E, \label{evephifbfb}
	\end{align}
	where, $\kf_E := \ket{\ebi}_E + \kebj_E$. So the result of Eve's attack on $\rho^{(0)}_{AT}$ is the following: 
	\begin{align*}
		\rho^{(1)}_{ATE} &= \half  \kb{0}_A \ot U\kbi_T U^\dagger + \half \kb{1}_A \ot U \kbphi_T U^\dagger \\
		%	&= \half  \kb{0}_A \ot P(\sum_b \ket{b, \ebi}_{TE}) + \half \kb{1}_A \ot P(\sqtt \sum_b \ket{b}_T \ot (\ket{\ebi}_E + \kebj_E)) \\
		&=\half  \kb{0}_A \ot P\left(\sum_{b = 0}^{D - 1} \ket{b, \ebi}_{TE}\right) + \half \kb{1}_A \ot P\left(\sqtt \sum_{b = 0}^{D - 1} \ket{b}_T \ot \kf_E\right),	
	\end{align*}
	where, recall, $P(\ket{z}) = \kb{z}$ is the projection operator. Henceforth, we will forgo writing the subscript for a register when the context is clear. Now, after the qudit arrives at Bob's lab, he measures the transit register $T$ in either POVM $Z$ or $X$ with equal probability. Let's consider the case when he uses $X$ and gets the outcome $\mbi - \kbphi$ (we are conditioning on a successful key-round for this analysis). This is the case when Bob sets his key-bit to $0$, because in a noiseless scenario, this outcome could only be obtained when Alice would have sent an $\ki$. Let's define the measurement operator in this case as $M_0 = \mbi_A \ot (\mbi - \kbphi) \ot \mbi_E $. Then the un-normalized post-measurement state, conditioned on him observing $M_0$ (again, we are only interested, for the moment, in a successful key distillation round) is:
	\begin{align}
		\rho_{ATE}^X &=  M_0 \cdot \rho^{(1)}_{ATE} \cdot M_0^\dagger \nonumber\\
		&= \half  \kb{0} \ot P\Big( ((\mbi - \kbphi) \ot \mbi) \sum_b \ket{b, \ebi} \Big) \nonumber\\
		&+ \half \kb{1} \ot P\Big(\sqtt \big((\mbi - \kbphi) \ot \mbi \big) \sum_b \ket{b} \ot \kf \Big) \nonumber\\	
		&= \half  \kb{0} \ot P\Big( \sum_b \ket{b, \ebi} - \half (\ket{i, e_i^i} + \ket{i, e_j^i} + \ket{j, e_i^i} + \ket{j, e_j^i} )\Big) \nonumber\\
		&+ \half \kb{1} \ot P\Big(\sqtt \big( \sum_b \ket{b, \fb} - \half (\ket{i, f_i} + \ket{i, f_j} + \ket{j, f_i} + \ket{j, f_j} )\big)\Big) \nonumber\\
		%	&= \half  \kb{0} \ot P\Big( \sum_{\slim} \ket{b, \ebi} + \half \ket{i, e_i^i} - \half \ket{i, e_j^i} - \half \ket{j, e_i^i} + \half \ket{j, e_j^i} \Big) \nonumber\\
		%	&+ \half \kb{1} \ot P\Big(\sqtt \big(\sum_{\slim} \ket{b \fb} +  \half \ket{i, f_i} - \half \ket{i, f_j} -\half \ket{j, f_i} + \half \ket{j, f_j} \big)\Big) \nonumber\\
		&= \half  \kb{0} \ot P\Big( \sum_{\slim} \ket{b, \ebi} + \half (\ki \otimes (\ket{\eii} - \ket{\eji})) - \half (\kj \otimes (\ket{\eii} - \ket{\eji}))\Big) \nonumber\\
		&+ \half \kb{1} \ot P\Big(\sqtt \big(\sum_{\slim} \ket{b, \fb} +  \half (\ket{i} \ot (\ket{f_i} - \ket{ f_j})) -\half (\ket{j} \ot (\ket{f_i} - \ket{ f_j}))\big)\Big) \nonumber\\
		&= \half  \kb{0} \ot P\Big( \sum_{\slim} \ket{b, \ebi} + \half \ket{i, g} -\half  \ket{j, g} \Big) \nonumber\\
		&+ \half \kb{1} \ot P\Big(\sqtt \big(\sum_{\slim} \ket{b, \fb} +  \half \ket{i, h} -\half \ket{j, h } \big) \Big), \label{eq:rhoxatewproj} 
	\end{align}	
	where in the last equality, we have defined $\ket{g} = \keii - \keji$ and $\ket{h} = \ket{f_i} - \ket{ f_j}$. 
	Now that Bob has his $X$ basis measurement result at his hand, we can trace out the transit register $T$ and add Bob's register $B$ to hold his measurement result. Then the resulting state is:  
	\begin{align*}
		\rho_{AEB}^X &= \half \kb{0}_A \ot \Big( \sum_{\slim} \kb{\ebi} + \frac{1}{2} \kb{g} \Big) \ot \kb{0}_B \\
		&+ \half \kb{1}_A \ot \half \Big(\sum_{\slim} \kb{\fb} + \half \kb{h} \Big) \ot \kb{0}_B.
	\end{align*}
	Similarly, if he uses POVM $Z$ and gets outcome $\mbi - \kbi$, he can be certain in a noiseless scenario that Alice has sent a $\kphi$. With the measurement operator $M_1 := \mbi_A \ot (\mbi - \kbi )  \ot \mbi_E$, in this case we get the following un-normalized post-measurement state:
	\begin{align*}
		\rho_{ATE}^Z &= M_1 \cdot  \rho^{(1)}_{ATE}  \cdot M_1^\dagger  \\
		&= \half  \kb{0} \ot P\Big( (\mbi - \kbi) \sum_b \ket{b, \ebi} \Big) + \half \kb{1} \ot P\Big(\sqtt (\mbi - \kbi) \sum_b \ket{b} \ot \kf\Big) \\
		&= \half  \kb{0} \ot P\Big( \smbni \ket{b, \ebi}\Big) + \half \kb{1} \ot P\Big(\sqtt \smbni \ket{b} \ot \kf\Big).
	\end{align*}
	Following a similar procedure as before, we trace out the transit register $T$ and add Bob's register holding his measurement result. The resulting density operator is:
	\begin{align*}
		\rho_{AEB}^Z &= \half \kb{0}_A \ot \smbni \kb{\ebi} \ot \kb{1}_B + \half \kb{1}_A \ot \half \smbni \kb{\fb} \ot \kb{1}_B.
	\end{align*}
	Then the total (still non-normalized) density operator that represents a key-bit generation round, is the following: 
	\begin{align}
		\rho_{AEB} &= \rho_{AEB}^X + \rho_{AEB}^Z \nonumber\\
		&= \half  \kb{0}_A \ot \Big( \sum_{\slim} \kb{\ebi} + \frac{1}{2} \kb{g} \Big) \ot \kb{0}_B \nonumber \\
		&+ \half \kb{1}_A \ot \half \Big(\sum_{\slim} \kb{\fb} + \half \kb{h}\Big) \ot \kb{0}_B \nonumber\\
		&+\half  \kb{0}_A \ot \smbni \kb{\ebi} \ot \kb{1}_B + \half \kb{1}_A \ot \half \smbni \kb{\fb} \ot \kb{1}_B \nonumber\\
		&= \half  \kb{0}_A \ot  \Big( \big( \sum_{\slim} \kb{\ebi} + \frac{1}{2} \kb{g}\big) \ot \kb{0}_B +  \smbni \kb{\ebi} \ot \kb{1}_B \Big) \nonumber\\
		&+ \half  \kb{1}_A \ot  \Big( \half \big(\sum_{\slim} \kb{\fb} + \half \kb{h} \big) \ot \kb{0}_B + \half \smbni \kb{\fb} \ot \kb{1}_B \Big) \label{eq:longdensityabe}.
	\end{align}
	Keeping in mind that, our ultimate goal is to bound Eve's entropy about Alice's register, i.e. $S(A|E)$, in the case where Alice and Bob shares a key-bit, we trace out Bob's register too, keeping only the registers of Alice and Eve. Thus, we calculate the final required density operator as:
	\begin{align}
		N\cdot\rho_{AE} &= \half  \kb{0}_A \ot  \Big(  \sum_{\slim} \kb{\ebi} + \frac{1}{2} \kb{g} +  \smbni \kb{\ebi} \Big) \nonumber\\
		&+ \half  \kb{1}_A \ot  \Big( \half \sum_{\slim} \kb{\fb} + \quart \kb{h} + \half \smbni \kb{\fb} \Big) \nonumber \\
		%	&= \half  \kb{0}_A \ot  \Big( \sum_{\slim} \kb{\ebi} + \frac{1}{2} \kb{g} +  \sum_{\slim} \kb{\ebi} + \kb{\eji} \Big) \nonumber\\
		%	&+ \half  \kb{1}_A \ot  \Big( \half \sum_{\slim} \kb{\fb} + \quart \kb{h} + \half \sum_{\slim} \kb{\fb} + \half \kb{\fj} \Big) \nonumber\\
		&= \kb{0}_A \ot  \Big( \half \sum_{\slim} \kb{\ebi} + \quart \kb{g} + \half  \sum_{\slim} \kb{\ebi} + \half \kb{\eji} \Big) \nonumber\\
		&+  \kb{1}_A \ot  \Big( \quart \sum_{\slim} \kb{\fb} + \frac{1}{8} \kb{h} + \quart \sum_{\slim} \kb{\fb} + \quart \kb{\fj} \Big) \nonumber\\
		&=  \kb{0} \ot \Big( \sum_{\slim} \kb{\ebi} + \frac{1}{2} \kb{\eji} + \quart\kb{g} \Big) \nonumber\\
		&+\kb{1} \ot \Big(\half \sum_{\slim} \kb{\fb} + \quart \kb{f_j} + \frac{1}{8} \kb{h} \Big) \label{eq:finaldensity},
	\end{align}
	where the normalization term $N$ can be calculated as:
	\begin{align}
		N &=\sum_{\slim}\bk{\ebi} + \half \bk{\eji} + \quart \bk{g} + \half \sum_{\slim} \bk{\fb} + \quart \bk{f_j} + \frac{1}{8} \bk{h} \label{eq:normalizerN}
	\end{align}
	
	Now, using Theorem \ref{theorem1}, we may compute the conditional entropy as:
	\begin{align}
		S(A|E) \ge  \sum_{\slim} \left(\frac{n_{b}^0 + n_{b}^1 }{N}\right) s_b + \left(\frac{n_{i}^0 + n_{i}^1 }{N}\right) s_{i} + (\frac{n_j^0 + n_j^1 }{N})s_j,\label{eq:entropy}
	\end{align}
	where:	
	\begin{align*}
		&n_b^0:= \bk{\ebi}, \qquad n_b^1 := \half \bk{f_b}, \text{ for all $b \ne i, j$}\\
		&n_{i}^0 := \half \bk{\eji}, \qquad n_{i}^1 := \frac{1}{8}\bk{h} , \\
		&n_j^0 := \quart \bk{g} , \qquad n_j^1 := \quart \bk{f_j}.
	\end{align*}
	
	and:
	\begin{align*}
		s_b &= H_2\left(\frac{n_b^0 }{n_b^0 + n_b^1}\right) - H_2\left(\half + \frac{\sqrt{(n_b^0 - n_b^1)^2 + 4 \times \Re^2 \bk{\ebi}{\sqtt \fb}} }{2(n_b^0 + n_b^1)} \right) \\
		s_{i} &= H_2\left(\frac{n_{i}^0 }{n_{i}^0 + n_{i}^1}\right) - H_2\left(\half + \frac{\sqrt{(n_{i}^0 - n_{i}^1)^2 + 4 \times \Re^2 \bk{\half \eji}{\frac{1}{2\sqrt{2}}h}} }{2(n_{i}^0 + n_{i}^1)} \right) \\
		s_j &= H_2\left(\frac{n_j^0 }{n_j^0 + n_j^1}\right) - H_2\left(\half + \frac{\sqrt{(n_j^0 - n_j^1)^2 + 4 \times  \Re^2 \bk{\half g}{\half f_j}} }{2(n_j^0 + n_j^1)} \right).
	\end{align*}
	
	Thus, to find a lower bound on $S(A|E)$ (thus giving us a lower bound on the protocol's key-rate), we must determine bounds for the inner-products appearing in the above expressions.  Of course, these inner products are functions of Eve's quantum ancilla.  We show how to determine suitable bounds on these systems based only on parameters that are directly observable in our protocol during test rounds (see Table \ref{tab:observable_stats}).
	
	\subsection{Parameter Estimation}
	%% ... which walks reader thru parameters
	To calculate the conditional entropy of $\rho_{AE}$, we need to estimate all the inner products appearing in equation \eqref{eq:normalizerN}. This can be done by connecting these inner products with observable noise statistics that arise from test rounds of Alice and Bob's quantum communication. 
	%	For example, the probability that Alice sends an $\ki$ and Bob measures a $\mbi - \kbphi$
	Let us see the statistics that can be observed directly in a test round. For example, in a round where Alice sends an $\ki$ or a $\kj$, Eve attacks with $U$, and Bob measures in $Z$; the probability that Bob gets a particular outcome $\kb{b}$ in $Z$ can be used to estimate partial $Z$ basis channel noise. In the following, by $p_{ib}$ and $p_{jb}$, we denote the probability that, given that Alice prepares $\ket{i}$ or $\ket{j}$ and Bob measures  the transit register in basis $Z$ then  the outcome is $\kb{b}$ for a particular $b \in \{0,... , D - 1\}$. (See also Table \ref{tab:observable_stats}.)
	\begin{align}
		\pib &= \bra{i} U^\dagger (\kb{b} \otimes \mbi) U \ki = \bk{\ebi} \label{eq:pib}\\
		\pjb &= \bra{j} U^\dagger (\kb{b} \otimes \mbi) U \kj = \bk{\ebj} \label{eq:pjb}
	\end{align}
	thus giving us $\{n_b^0\}$ as needed in the entropy equation.
	Now, we trace the evolution of the quantum system when Alice prepares $\ki$ and Bob measures in POVM $X$. For example, we can not observe the inner product $\bk{\eii}{\eji}$ directly. But we consider the probability that Alice sends an $\ki$, Eve attacks with $U$, and Bob measures in $X$ to find a $\kphi$, denoted as $\piphi$:	
	\begin{align}
		\piphi &= \bra{i}U^\dagger (\kb{\phi} \otimes \mbi) U\ki \nonumber\\
		&= \sum_{b, c} \bra{b} \kb{\phi} \ket{c} \bk{\ebi}{e_c^i}  \nonumber\\
		&= \half \sum_{b, c}  \bra{b} (\kb{i} + \kb{i}{j} + \kb{j}{i} + \kbj) \ket{c} \bk{\ebi}{e_c^i} \nonumber\\
		&=\half (\bk{\eii} + \bk{\eii}{\eji} + \bk{\eji}{\eii} + \bk{\eji}) \nonumber\\
		&= \half (\pii + 2 \Re \bk{\eii}{\eji} + \pij), \label{eq:piphito2re}
	\end{align}
	where, we have used equation \eqref{eq:pib} for $\pii, \pij$ and an elementary property of complex inner products. Notice that, even though we could not observe $\bk{\eii}{\eji}$, equation \eqref{eq:piphito2re} will imply:
	\begin{align}
		&2 \Re \bk{\eii}{\eji} = 2\piphi - \pii - \pij \label{eqreiieji}. 
	\end{align}
	Using this estimation of $\Re \bk{\eii}{\eji}$, we can now estimate the inner product $\bk{g}$, which appears in the normalizer $N$ in equation \eqref{eq:normalizerN}. This is:
	\begin{align}
		\bk{g} &= (\bra{\eii} - \bra{\eji})(\ket{\eii} - \ket{\eji}) \nonumber\\
		&= \bk{\eii} - \bk{\eii}{\eji} - \bk{\eji}{\eii} + \bk{\eji} \nonumber \\ 
		&= \pii - 2 \Re \bk{\eii}{\eji} + \pij \nonumber\\
		&= 2\pii + 2\pij - 2 \piphi. 
	\end{align}
	Now let's focus on calculating $\bk{f_b}$. Remembering that $\kf = \ket{\ebi} + \kebj$ (See equation \eqref{evephifbfb}),  we can easily derive the following:
	\begin{align}
		\bk{f_b} &= (\bra{\ebi} + \bra{\ebj})(\ket{\ebi} + \kebj) \nonumber\\
		&= \bk{\ebi} + \bk{\ebi}{\ebj} + \bk{\ebj}{\ebi} + \bk{\ebj} \nonumber\\			
		& = \pib + 2 \Re \bk{\ebi}{\ebj} + p_{jb}, \label{eq:fbfbtoebiebj}
	\end{align}
	where we have used equation \eqref{eq:pib} and \eqref{eq:pjb} for $\pib, \pjb$. Now if we look closely at equation $\eqref{evephifbfb}$, we see that $\bk{f_b}$ is actually directly observable. Because it is the probability that Alice sends a $\kphi$, Eve attacks with $U$, and conditioned on the case that Bob measures in $Z$, gets an outcome $\ket{b}$. We denote it by $p_{\phi b}$ and see that:
	\begin{align}
		p_{\phi b} &= \bra{\phi}U^\dagger (\kb{b} \otimes \mbi ) U\kphi \nonumber\\
		&= \Big(\sqtt \sum_{c = 0}^{D - 1} \bra{c} \ot \bra{f_c} \Big) (\kb{b} \otimes \mbi ) \Big(\sqtt \sum_{d = 0}^{D - 1} \ket{d} \ot \ket{f_d} \Big) \nonumber\\
		&=\half \sum_{c,d= 0}^{D - 1} \bra{c} \kb{b} \ket{d} \bk{f_c}{f_d} \nonumber\\
		&=\half \bk{f_b}, \label{eq:phibfbfbdirect}
	\end{align}  
	and consequently, $\bk{f_b} = 2 \pphib$. So, from equations \eqref{eq:fbfbtoebiebj} and \eqref{eq:phibfbfbdirect} we infer the following:
	\begin{align}
		2 \Re \bk{\ebi}{\ebj} = 2 \pphib - \pib - p_{jb}. \label{eq:anybfbfb} 
	\end{align}
	Notice that equation \eqref{eq:fbfbtoebiebj} and consequently \eqref{eq:anybfbfb}, holds for all $b = 0, ..., D - 1$. So we immediately get $\bk{\fj}{\fj}$ for normalizer $N$.
	%	\begin{align}
	%	\bk{f_j} &= \pij + 2 \Re \bk{\eji}{\ejj} + p_{jj} 		\implies 2 \Re \bk{\eji}{\ejj} = 2 p_{\phi j}  - \pij - \pjj  \label{eq:rjijj}
	%	\end{align}
	%		\begin{align}
	%		\bk{f_i}{f_j} &= \bk{\eii}{\eji} + \bk{\eii}{\ejj} + \bk{\eij}{\eji} + \bk{\eij}{\ejj} \\
	%		\bk{f_j}{f_i} &= \bk{\eji}{\eii} + \bk{\eji}{\eij} + \bk{\ejj}{\eii} + \bk{\ejj}{\eij}
	%		\end{align}
	Now let's calculate the last inner product in $N$ which is $\bk{h}$. First let's discover the constituent inner products for $\bk{h}$. Then we will connect each of those to Alice and Bob's observables. 
	\begin{align}
		\bk{h} &= (\bfi - \bfj)(\kfi - \kfj) \nonumber\\
		&= \bk{f_i} - \bk{f_i}{f_j} - \bk{f_j}{f_i} + \bk{f_j} \nonumber\\
		&= \bk{f_i} -2 \Re \bk{f_i}{f_j} + \bk{f_j}. \label{eq:hh}
	\end{align}
	Now, let us take advantage of another directly observable quantity. Which is the probability that Bob would measure a $\kphi$ in the $X$ basis if Alice indeed sent a $\kphi$. We denote it as $p_{\phi \phi}$ and see that:
	\begin{align}
		p_{\phi \phi} &= \bra{\phi} U^\dagger (\kbphi \otimes \mbi) U \kphi  \nonumber\\
		&= \sqtt \Big(\sum_{b} \bra{b, f_b}\Big) \big(\kbphi \otimes \mbi\big) \sqtt \Big(\sum_{c} \ket{c, f_c}\Big)\nonumber \\
		&= \quart \Big( \sum_{b, c} \big(\bra{b} \kb{i} \ket{c} \otimes \bk{f_b}{f_c} + \bra{b} \kb{i}{j} \ket{c}  \otimes \bk{f_b}{f_c}  \nonumber\\
		&+ \bra{b} \kb{j}{i} \ket{c}  \otimes \bk{f_b}{f_c} + \bra{b} \kb{j} \ket{c}  \otimes \bk{f_b}{f_c} \big)\Big) \nonumber \\
		&= \quart (\bk{f_i}+\bk{f_i}{f_j} + \bk{f_j}{f_i} +\bk{f_j}) \nonumber\\
		&= \quart (\bk{f_i}+ 2 \Re \bk{f_i}{f_j} +\bk{f_j}) \nonumber\\				
		&= \half (p_{\phi i} + \Re \bk{f_i}{f_j} + p_{\phi j}). \label{eq:pphiphitofifj} 
	\end{align}
	Equation \eqref{eq:pphiphitofifj} implies that:
	\begin{align}
		\Re \bk{f_i}{f_j} = 2 p_{\phi \phi} - p_{\phi i} - p_{\phi j}. \label{eq:rfifj}
	\end{align}
	Along with the fact that $\bk{f_i} = 2 p_{\phi i}$ and $\bk{f_j} = 2 p_{\phi j}$ from equation \eqref{eq:phibfbfbdirect}, we can say from equation \eqref{eq:hh} and \eqref{eq:rfifj} that:
	\begin{align*}
		\bk{h} &= 4(p_{\phi i} + p_{\phi j} - p_{\phi \phi}). 
	\end{align*}
	This concludes the estimation of inner products appearing in the normalizer $N$ of $\rho_{AE}$ in \eqref{eq:normalizerN}. We need to estimate the real parts of three more classes of inner products to calculate each of the $\lambda$-terms appearing in Theorem \ref{theorem1}. These are $\{\Re\bk{\ebi}{\sqtt \fb}\}_b,\Re \bk{\half g}{\half f_j},\Re \bk{\half \eji}{\frac{1}{2\sqrt{2}} h}$. In the following we connect these inner products to observable statistics. Let's focus on the first one:
	\begin{align} 
		\Re\bk{\ebi}{\sqtt \fb} &=\sqtt  \Re (\bra{\ebi}) (\ket{\ebi} + \ket{\ebj}) \nonumber\\
		&=  \sqtt \Re (\bk{\ebi} + \bk{\ebi}{\ebj}) \nonumber\\
		&= \sqtt (\pib + \pphib -\frac{\pib}{2} - \frac{\pjb}{2}) \label{fct}, % first cross term 
	\end{align}		
	where we have used the definition of $\ket{ f_b}$ in the first equality and have used equation \eqref{eq:pib} and \eqref{eq:anybfbfb} to insert the value of  $\Re \bk{\ebi} , \Re \bk{\ebi}{\ebj}$. Now let's focus on the second inner product $\Re \bk{\half g}{\half f_j}$:
	\begin{align}
		\Re \bk{\half g}{\half f_j} &= \quart  \Re (\bra{\eii} - \bra{\eji})(\ket{\eji} + \ket{\ejj}) \nonumber\\
		&= \quart \Re(\bk{\eii}{\eji} + \bk{\eii}{\ejj} - \bk{\eji}{\eji} - \bk{\eji}{\ejj}) \nonumber \\
		&= \quart(\piphi - \frac{\pii}{2} + \Re\bk{\eii}{\ejj} - \pij - \pphij + \frac{\pjj}{2}). \label{sct} % second cross term.
	\end{align}
	The value of $\Re \bk{\eii}{\eji}$ and $\Re \bk{\eji}{\ejj}$ is found in $\eqref{eqreiieji}$ and \eqref{eq:anybfbfb} respectively.
	Furthermore, $\bk{\eji}{\eji}$ is simply $\pij$ because of equation \eqref{eq:pib}. Noticeably, the term $\bk{\eii}{\ejj}$ is not observable. We will deal with this a bit later. Now we move on to the last of the necessary inner products for the theorem, $\Re \bk{\half \eji}{\frac{1}{2\sqrt{2}} h}$. 
	\begin{align}
		\half \times  \frac{1}{2\sqrt{2}} \Re \bk{\eji}{h}&= \quart \Re \bra{\eji}(\ket{f_i} - \ket{f_j}) \nonumber\\ 
		&= \quart \Re \bra{\eji}( \ket{\eii} + \ket{\eij} - \ket{\eji} - \ket{\ejj}) \nonumber\\
		&= \quart \Re (\bk{\eji}{\eii} + \bk{\eji}{\eij} - \bk{\eji}{\eji} - \bk{\eji}{\ejj}) \nonumber\\
		&= \quart\left(\piphi - \frac{\pii}{2} + \Re\bk{\eji}{\eij} - \pij - \pphij + \frac{\pjj}{2}\right), \label{tct} % third cross term.
	\end{align}
	where the unknown terms $\Re \bk{\eji}{\eii} = \Re \bk{\eii}{\eji}, \Re \bk{\eji}{\ejj}$ can be found in equations \eqref{eqreiieji} and \eqref{eq:anybfbfb} respectively. In equation \eqref{tct}, we are again faced with a term $\Re \bk{\eji}{\eij}$ for which we do not have a direct observation. Now we deal with this term and the other unobservable term from equation \eqref{sct}, which is $\Re \bk{\eii}{\ejj}$. Although it is not hard to see, we need to take several steps to find an equation that relates these two inner products.
	%	\begin{align*}
	%	\bk{f_i}{\fj} + \bk{\fj}{f_i} = 2 \Re \bkfifj.
	%	\end{align*}
	Consider $\bk{f_i}{\fj}$ which we may expand as:
	\begin{align}
		2 \Re \bkfifj &= 2 \Re(\bk{\eii}{\eji} + \bk{\eii}{\ejj} + \bk{\eij}{\eji} + \bk{\eij}{\ejj}) \label{eq:refifj}
	\end{align}
	First let's deal with the unobservable term $\bk{\eij}{\ejj}$. It is easy to see that, the probability of Alice sending a $\kj$, Eve attacking with $U$ and Bob measures in $X$ to find a $\kphi$, denoted by $p_{j\phi}$ is:
	\begin{align}
		\pjphi &= \bra{j}U^\dagger (\kb{\phi} \otimes \mbi) U\kj \nonumber\\
		&= \sum_{b, c} \bra{b} \kb{\phi} \ket{c} \bk{\ebj}{e_c^j}  \nonumber\\
		&= \half \sum_{b, c}  \bra{b} (\kb{i} + \kb{i}{j} + \kb{j}{i} + \kbj) \ket{c} \bk{\ebj}{e_c^j} \nonumber\\
		&=\half (\bk{\eij} + \bk{\eij}{\ejj} + \bk{\ejj}{\eij} + \bk{\ejj}) \nonumber\\
		&= \half (\pji + 2 \Re \bk{\eij}{\ejj} + \pjj). \label{eq:pjphito2re}
	\end{align} 
	From equation \eqref{eq:pjphito2re}, it is clear that:
	\begin{align}
		2 \Re \bk{\eij}{\ejj} = 2 p_{j\phi} - \pji - \pjj. \label{eq:reeijjj}
	\end{align}
	The value of one of the other three unobservable terms appearing in equation \eqref{eq:refifj}, i.e., $\bk{\eii}{\eji}$ can be found in equation \eqref{eqreiieji}. 
	However,  $\Re \bk{\eii}{\ejj}, \Re \bk{\eij}{\eji})$ are unobservable at this point. With the help of equations \eqref{eqreiieji} and \eqref{eq:reeijjj}, we can rewrite equation \eqref{eq:refifj} as:
	\begin{align}
		2 \Re \bkfifj &= 2\piphi  - \pii - \pij  + 2 \Re (\bk{\eii}{\ejj} + \bk{\eij}{\eji}) + 2\pjphi - \pji - \pjj, \label{eq:unobeq1} 
	\end{align}
	We further notice from equation $\eqref{eq:rfifj}$, 
	\begin{align}
		2\Re \bkfifj = 4 p_{\phi \phi} - 2p_{\phi i} - 2p_{\phi j}.\label{eq:unobeq2} 
	\end{align}
	Then, we equate the previous two equations \eqref{eq:unobeq1}, \eqref{eq:unobeq2} to ultimately find:
	\begin{align}
		4 p_{\phi \phi} - 2p_{\phi i} - 2p_{\phi j} &= 2\piphi  - \pii - \pij  + 2 \Re (\bk{\eii}{\ejj} + \bk{\eij}{\eji}) + 2\pjphi - \pji - \pjj  \nonumber\\
		\implies 2\Re (\bk{\eii}{\ejj} + \bk{\eij}{\eji}) &= 4 p_{\phi \phi} - 2p_{\phi i} - 2p_{\phi j} - 2\piphi +  \pii + \pij - 2 \pjphi + \pji + \pjj  \nonumber\\
		\implies \Re (\bk{\eii}{\ejj} + \bk{\eij}{\eji}) &= 2 p_{\phi \phi} - p_{\phi i} - p_{\phi j} - \piphi +  \frac{\pii}{2} + \frac{\pij}{2} - \pjphi + \frac{\pji}{2} + \frac{\pjj}{2} \label{sm2unobs}. 
	\end{align}
	Let's define the right-hand side of equation \eqref{sm2unobs} to be $K$, the value of which may be computed by Alice and Bob based only on observed statistics of the quantum channel.  Now we have all the pieces necessary to compute the conditional entropy $S(A|E)$ according to Theorem \ref{theorem1}. We minimize $S(A|E)$ given by Equation \ref{eq:entropy} over the two unobservable values.  Note that we minimize over these unobservable quantities as we must assume that Eve choose the attack strategy that gives her the most information.  However, her attack must be constrained by the above analysis.  These unobservable inner-products are further constrained by Cauchy-Schwarz in that $-\sqrt{\pij\cdot\pji} \le Re\bk{\eji}{\eij} \le \sqrt{\pij\cdot \pji}$ and similarly for the other inner product.  The sum of these two values is further constrained by Equation \ref{sm2unobs} (the value $K$). With the bound on $S(A|E)$ calculated, we can focus on the Shannon entropy of Alice's register given Bob's register, i.e. $H(A|B)$. To see it clearly, let us remember the density operator $\rho_{AEB}$ in equation \eqref{eq:longdensityabe}, where Bob's register was still in place. From that $\rho_{AEB}$, we can see that the probability of Alice and Bob sharing different pairs of classical bits $0$ and $1$ is the following:
	\begin{align*}
		p_{00} &= \frac{1}{2N} \Big(  \sum_{\slim} \bk{\ebi} + \half \bk{g}\Big) = \frac{1}{2N} \Big(\sum_{\slim} \pib +   \pii +   \pij -   \piphi\Big) \\
		p_{01} &= \frac{1}{2N} \smbni \kb{\ebi} = \frac{1}{2N}\Big(\sum_{\slim} \pib + \pij\Big) \\
		p_{10} &= \frac{1}{2N} \Big(\half \sum_{\slim} \kb{\fb} + \quart \kb{h} \Big) = \frac{1}{2N}\Big(  \sum_{\slim} \pphib + (\pphii + \pphij - \pphiphi)\Big)\\
		p_{11} &=\frac{1}{4N}\smbni \kb{\fb} = \frac{1}{2N} \Big(  \sum_{\slim} \pphib +   \pphij\Big).
	\end{align*}
	From $p_{00}, p_{01}, p_{10}, p_{11}$, it is trivial to compute $H(A|B)$. From all of this, we may easily compute a lower-bound on the min-entropy $S(A|E)$ and also directly compute $H(A|B)$ thus giving us a lower-bound on the key-rate of this protocol.
	
	\subsection{Evaluation}
	%% Compare both versions (|i>+|j> and equal superpos) to HD-BB84 under a variety of channels
	
	Note that the above security analysis and bound of $S(A | E)$ and $H(A|B)$, would hold for an arbitrary quantum channel; one need only observe those values listed in Table \ref{tab:observable_stats} in order to minimize $S(A|E)$ as described in the previous section. As examples, and to compare with other protocols, we will evaluate our protocol in two different channels, commonly used in QKD protocol evaluation.  These are the depolarizing channel and the amplitude damping channel. First, let's consider the depolarization channel.   Given a density operator $\sigma$ acting on a Hilbert space of dimension $D$, the depolarization channel with parameter $Q$, denoted here as $\mathcal{E}_Q$ acts as follows:
	\begin{equation}
		\mathcal{E}_Q(\sigma) = \left(1-\frac{D}{D-1}Q\right)\sigma + \frac{Q}{D-1}I. 
	\end{equation}
	To calculate the key rate of our protocol, we calculate the required observable statistics assuming the adversary uses this channel (in particular, the statistics indexed in Table \ref{tab:observable_stats}).  These are easily found to be:
	\begin{align*}
		\pii &= \pjj = \pphiphi = 1 - Q \\
		\pib &= \pjb = \pphib =  \frac{Q}{D - 1} \\
		\piphi = \pjphi = \pphii &= \pphij = \half \Big(1 - \frac{DQ}{D - 1}\Big) + \frac{Q}{D - 1}. 
	\end{align*} 
	
\begin{figure}
		\centering
		\includegraphics[width=0.8\linewidth]{"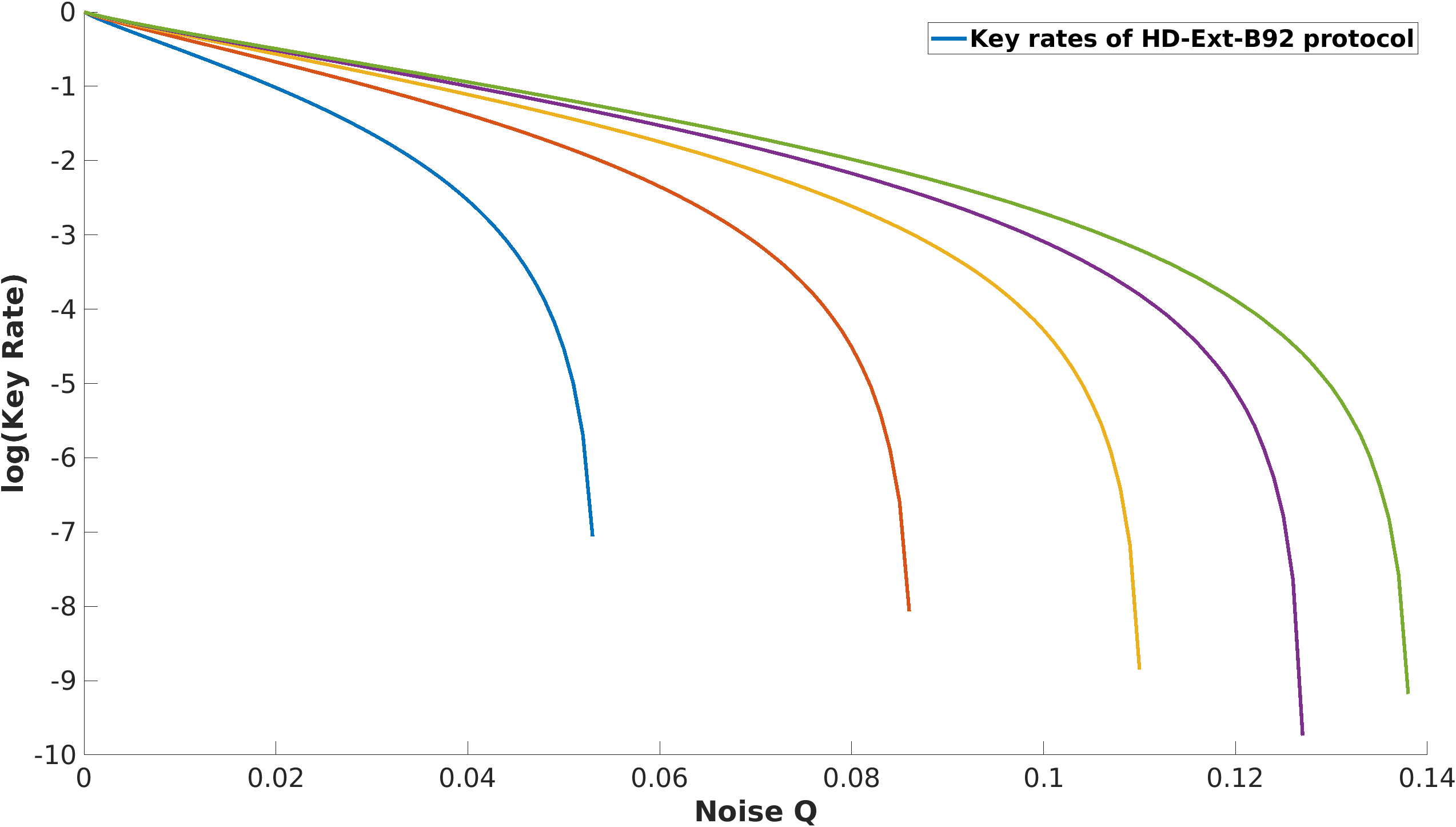"}
		\caption{The key-rate of the HD-Ext-B92 protocol for various dimensions $D$ assuming a depolarization channel.  Here the dimension increases from left-to-right from $D=2^1$ to $D=2^5$ in powers of two. We observe numerically, as the dimension continues to increase, the noise tolerance for this channel tends towards $15.5\%$.}
		\label{fig:our-prot-only}
	\end{figure}

	This is sufficient to evaluate the key-rate of our protocol as shown in Figure \ref{fig:our-prot-only}.  Note that, as with other high-dimensional QKD protocols, as the dimension of the system increases, the tolerance to depolarization noise also increases.  In our numerical evaluations, the noise tolerance approaches $15.5\%$ as the dimension increases thus showing that, as with other high-dimensional QKD protocols, the extended-B92 style scheme can also benefit from higher dimensional systems, at least against this particular channel type.
	
\begin{figure}
		\centering
		\includegraphics[width=0.8\linewidth]{"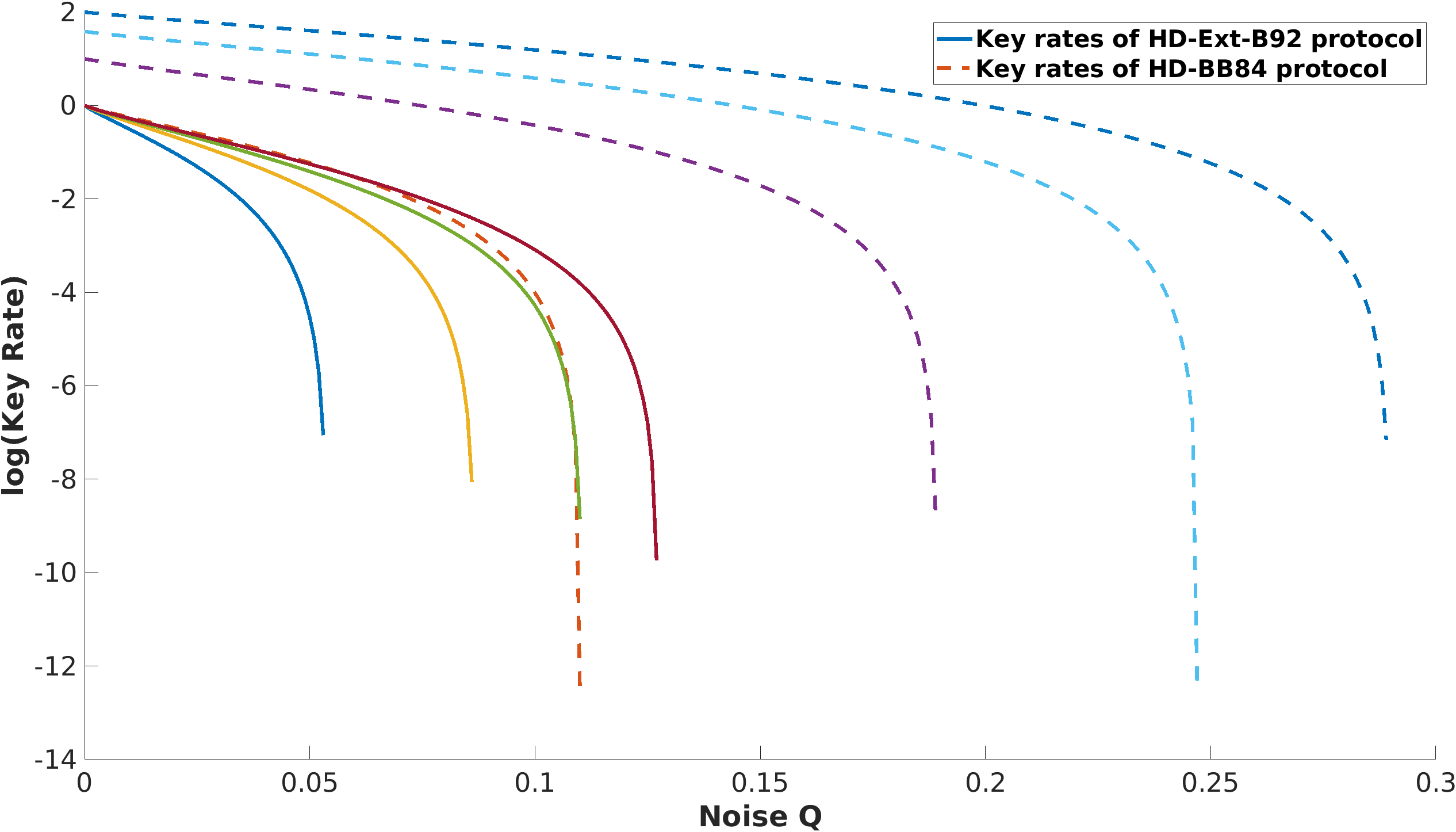"}
		\caption{Key-rate comparison between our HD-Ext-B92 protocol (solid lines) and the HD-BB84 protocol (dashed lines).  Here the dimension for each protocol line individually increases left-to-right from $D=2^1$ up to $D=2^4$.  Note that for $D=2^1$, HD-BB84 has a noise tolerance of $11\%$ (as expected since, in that case, it is standard BB84) while the HD-Ext-B92 protocol does not attain that level of noise tolerance until $D=2^3$.  See text for further discussion.}
		\label{fig:hddc-ours-bb84}
	\end{figure}

	We also compare with the HD-BB84 protocol \cite{cerf2002security} which we now state for completeness.  Similar to the qubit case, the qudit based HD-BB84 uses two bases, namely, the computational basis $Z  = \{ \ket{0}, \ket{1}, ..., \ket{D - 1} \}$ and another basis denoted by $X$ where $X = \{\ket{x_0}, \ket{x_1},...,\ket{x_f} \}$.  We assume the two bases are mutually unbiased. Alice sends basis states from these two bases and Bob measures in $X$ or $Z$.  If both parties chose the $Z$ basis, the result is used for their raw key; otherwise, if both parties choose the $X$ basis, they use this to measure the noise in the quantum channel. The unconditional security of this protocol has been proven \cite{QKD-BB84-rate1, koashi2009simple}. An entropic uncertainty relation presented in \cite{berta2010uncertainty} can be used to easily derive the following asymptotic key rate $r$ for HD-BB84 assuming a depolarization channel with a noise parameter $Q$. The final equation reads:
	\begin{align*}
		r = \log D - 2(H_2 (Q) + Q \log(D - 1)),
	\end{align*}
	The result of this comparison is presented in Figure \ref{fig:hddc-ours-bb84}. As expected, BB84 outperforms our protocol.  However, this is not surprising as BB84 at the qubit level also outperforms the B92 and Extended B92 protocol.  Furthermore, our high-dimensional protocol is not even utilizing two complete bases as HD-BB84 does; instead, we are using a weak version where Alice need only send three states and Bob need only perform partial measurements in the second basis.  Note also that we did not choose an optimal basis choice and, indeed, alternative encoding selections for the $X$ state may lead to higher key rates for our HD-Ext-B92 protocol as demonstrated at least for the qubit case \cite{lucamarini2009robust,QKD-Tom-Krawec-Arbitrary-2016}.

	Another channel of interest is the amplitude damping channel which can be described its Kraus operators: 
	\begin{equation}
		E_0 = \left(\begin{array}{ccccc}
			1 & 0 & 0 &\cdots & 0\\
			0 & \sqrt{{1-p}} & 0 &\cdots & 0\\
			0 & 0 & \sqrt{{1-p}} &\cdots & 0\\
			\vdots & \vdots & \vdots & \ddots & \vdots\\
			0 & 0 & 0 &\cdots & \sqrt{{1-p}}\end{array}\right)
	\end{equation}
	and:
	\begin{equation}
		E_1 = \left(\begin{array}{ccccc}
			0 & \sqrt{{p}} & 0 &\cdots & 0\\
			0 & 0 & 0 &\cdots & 0\\
			0 & 0 & 0 &\cdots & 0\\
			\vdots & \vdots & \vdots & \ddots & \vdots\\
			0 & 0 & 0 &\cdots & 0\end{array}\right),
		\cdots ,
		E_{D-1} = \left(\begin{array}{ccccc}
			0 & 0& 0 &\cdots & \sqrt{{p}} \\
			0 &0& 0 &\cdots &0\\
			0 & 0 & 0 &\cdots &0\\
			\vdots & \vdots & \vdots & \ddots & \vdots\\
			0 & 0 & 0 &\cdots & 0\end{array}\right).
	\end{equation}

	\begin{table}[ht]
		\centering
		\caption{Key rates for high-dimensional extended B92 protocol with $D = 4$ and different choices of $\ki$ and $\kj$ under an amplitude damping channel with parameter $p = .08$.}
		\begin{tabular}[t]{|c|c|}
			\hline
			$\kphi = \frac{1}{\sqrt{2}} (\ki + \kj)$  &  key rate \\
			\hline
			$\ki = \ket{0}, \kj = \ket{1}$ & .85 \\
			$\ki = \ket{0}, \kj = \ket{2}$ & .85 \\
			$\ki = \ket{1}, \kj = \ket{2}$ & .29 \\
			$\ki = \ket{1}, \kj = \ket{3}$ & .29 \\		
			\hline
		\end{tabular}
		\label{tab:ampl_chan}
	\end{table}	
	As before, we can compute those observable parameters in Table \ref{tab:observable_stats} under this channel and then use our analysis in the previous section to derive a lower-bound on the key-rate of our protocol.  We can see that the key rate of our protocol can vary significantly with the choice of basis states, in particular the distinguished $\ket{i}$ and $\ket{j}$ as shown in Table \ref{tab:ampl_chan}.  Since these are set by the users, they may choose basis states based on the channel properties to maximize the key rate.
	
	\section{Closing Remarks}
	In this work, we have presented the usage of high-dimensional quantum systems as communication resources between Alice and Bob in the extended B92 protocol, originally introduced in \cite{lucamarini2009robust} for qubits. When extending that protocol to higher dimensions we took care to attempt to minimize the quantum resources used by parties.  In particular, our protocol only requires Alice to send three different states while Bob need only perform partial measurements.

We performed an information theoretic security analysis against collective attacks and evaluated under two different channels, the depolarization channel and the amplitude damping channel.  We showed that, as with other high-dimensional protocols, under a depolarization channel the noise tolerance tends to increase with the dimension of the system.  For the HD-Ext-B92 protocol, this tolerance eventually converges to 15.5\% (as observed by our numerical computations).  Under an amplitude damping channel, we showed how the choice of basis states used can greatly affect the key rate of the overall protocol.

Perhaps the biggest open question at the moment is to determine the effects of alternative superposition states on the protocol.  We only considered a state of the form $\frac{1}{\sqrt{2}}\ket{i} + \frac{1}{\sqrt{2}}\ket{j}$.  One obvious candidate to consider would be the effect of having Alice send an equal superposition state.  Unfortunately, the security analysis of such a protocol proved to be highly difficult, especially when using the technique of mismatched measurements (as we used here).  The analysis might be simplified by having Alice send complete basis states instead of only a small subset of basis states in which case alternative proof methods may be applied.  We leave this interesting question as future work.  We also only performed an asymptotic key rate analysis - performing a finite key analysis, taking into account also, perhaps, less ideal measurement devices, would also be interesting to consider.

	\bibliography{references}
	\bibliographystyle{unsrt}
\end{document}